\documentclass[twocolumn,showpacs,preprintnumbers,amsmath,amssymb,groupedaddress,pre]{revtex4}

\usepackage{graphicx}
\usepackage{dcolumn}
\usepackage{bm}
\usepackage{epsfig}

\def\e{\begin{equation}}
\def\f{\end{equation}}
\def\=#1{\overline{\overline #1}}

\def\-#1{{\bf #1}}
\def\.{\cdot}
\def\l#1{\label{eq:#1}}
\def\r#1{(\ref{eq:#1})}
\def\vec#1{{\bf #1}}

\begin{document}

\title{Boundary conditions for interfaces of electromagnetic (photonic) crystals and generalized Ewald-Oseen extinction principle}

\author{Pavel A. Belov}
\affiliation{Queen Mary College, University of London, Mile End
Road, London, E1 4NS, United Kingdom} \affiliation{Photonics and
Optoinformatics Department, St. Petersburg State University of
Information Technologies, Mechanics and Optics, Sablinskaya 14,
197101, St. Petersburg, Russia}


\author{Constantin R. Simovski}
\affiliation{Photonics and Optoinformatics Department, St. Petersburg State University of Information Technologies, Mechanics and Optics, Sablinskaya 14, 197101, St. Petersburg, Russia}

\date{\today}

\begin{abstract}
The problem of plane-wave diffraction on semi-infinite orthorhombic
electromagnetic (photonic) crystals of general kind is considered.
Boundary conditions are obtained in the form of infinite system of
equations relating amplitudes of incident wave, eigenmodes excited
in the crystal and scattered spatial harmonics. Generalized
Ewald-Oseen extinction principle is formulated on the base of
deduced boundary conditions. The knowledge of properties of infinite
crystal's eigenmodes provides option to solve the diffraction
problem for the corresponding semi-infinite crystal numerically. In
the case when the crystal is formed by small inclusions which can be
treated as point dipolar scatterers with fixed direction the problem
admits complete rigorous analytical solution. The amplitudes of
excited modes and scattered spatial harmonics are expressed in terms
of the wave vectors of the infinite crystal by closed-form
analytical formulae. The result is applied for study of reflection
properties of metamaterial formed by cubic lattice of split-ring
resonators.
\end{abstract}

\maketitle

\section{Introduction}

Electromagnetic crystals are artificial periodical structures
operating at the wavelengths comparable with their periods
\cite{Sakoda,PhotJMW,specialis}. At the optical frequencies such
structures are called as photonic crystals. The inherent feature of
these materials is the existence of frequency bands where the
crystal does not support propagating waves. The band gaps are caused
by spatial resonances of the crystal lattice and strongly depend on
the direction of propagation. It means that electromagnetic crystals
are media with spatial dispersion \cite{Agranovich,Agarwal,Birman}.
The material parameters: permittivity and permeability for such
materials, if they can be introduced at all, depend on the wave
vector as well as on the frequency. Notice, that the homogenization
approach is not the most convenient way for the description of
electromagnetic crystals even at low frequencies. It often requires
introduction of additional boundary conditions in order to describe
boundary problems correctly, and this involves related complexities.
The photonic and electromagnetic crystals are usually studied with
the help of numerical methods \cite{Sakoda,PhotJMW,specialis}.
Analytical models exist only for a very narrow class of the
crystals. Some types of the crystals can be studied analytically
under a certain approximation, but the strict analytical solution
for a photonic crystal is an exception.

The goal of the present paper is to demonstrate how boundary
problems for electromagnetic crystals can be effectively studied
using analytical methods. The paper is separated into the two parts.
In the first part the boundary conditions for electromagnetic
crystals of general kind are deduced in the form of infinite system
of equations relating amplitudes of incident wave, excited
eigenmodes of the crystal and scattered spatial harmonics. This
system can be interpreted as generalization of well-known
Ewald-Oseen extinction principle \cite{Ewald,Oseen,BornWolf} which
states that the polarization of dielectric is distributed so that it
cancels out the incident wave and produces the propagating wave. For
the electromagnetic crystals, inherently periodic structures, the
generalized Ewald-Oseen principle states that the polarization of
dielectric is distributed so that it cancels out the incident wave
as well as all spatial harmonics associated with periodicity of the
boundary. This principle expressed in the form of infinite system of
boundary conditions provides opportunity to solve the boundary
problem for semi-infinite crystal of certain kind numerically if the
eigenmode problem for corresponding infinite crystal is already
solved. In the second part of the paper the proposed approach is
applied for the case of electromagnetic crystals formed by small
inclusions which can be treated as point dipolar scatterers with
fixed direction. In this case the system of boundary conditions
admits complete rigorous analytical solution. The amplitudes of
excited eigenmodes and scattered spatial harmonics are expressed in
terms of wavevectors of eigenmodes using closed-form analytical
formulae. These results are unique extension and generalization of
known Mahan-Obermair theory \cite{Mahan} for the case then period of
the crystal is compared with wavelength. At the end of the paper it
is demonstrated how reflection from the semi-infinite cubic lattice
of resonant scatterers (split-ring resonators) can be modeled in the
regime of strong spatial dispersion observed in such crystals
\cite{Belovhomo}.

\section{Proof of generalized Ewald-Oseen extinction principle}

\begin{figure}[h]
\centering \epsfig{file=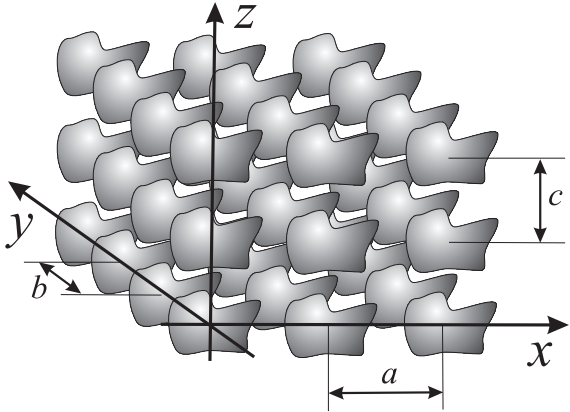, width=7.5cm} \caption{Geometry
of an infinite electromagnetic crystal} \label{gengeom}
\end{figure}

In this section we provide proof of generalized Ewald-Oseen
extinction principle for arbitrary semi-infinite electromagnetic
crystal with orthorhombic elementary cell. First, let us consider an
infinite orthorhombic electromagnetic crystal with geometry
schematically presented in Fig.~\ref{gengeom} and characterized by
three-periodical permittivity distribution: \e \=\varepsilon(\vec
r)=\=\varepsilon(\vec r+\vec a n+\vec b s+\vec c l). \f In this
expression and in the further text the two lines over a quantity
designate that the quantity is dyadic (tensor of second rank in
three-dimensional space). It means that we consider  of the most
general kind of electromagnetic crystals formed by dielectrics.

\begin{figure}[h]
\centering \epsfig{file=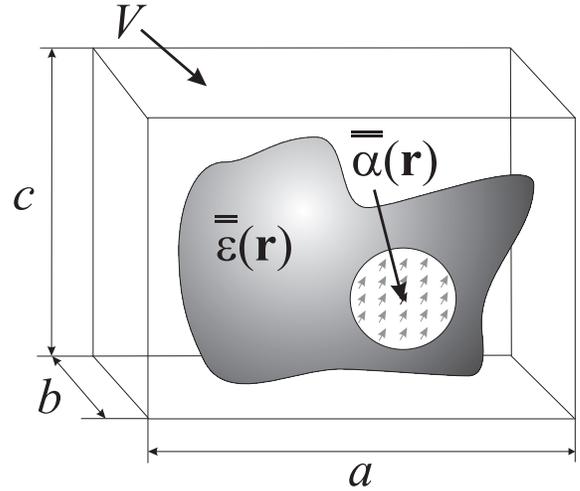, width=7.5cm}
\caption{Illustration for replacement of dielectric permittivity by
local polarizability} \label{local}
\end{figure}

In this paper we are using local field approach, an unconventional
method for description of fields inside dielectrics. We will operate
with local parameters like polarization density $\vec P$ and local
electrical field $\vec E_{\rm loc.}$, but not with average electric
field $\vec E$ and displacement $\vec D$ as usual. The similar
approach was used in \cite{BornWolf} for rigorous derivation of
Ewald-Oseen extinction theorem and in \cite{Birman}. The dielectric
can be treated as a very dense cubic lattice of point scatterers
with ceratin local polarizability. In this formulation the
dielectric permittivity $\=\varepsilon(\vec r)$ has to be replaced
(see Figure \ref{local} for illustration) by the local
polarizability $\=\alpha(\vec r)$ relating the bulk polarization
density $\vec P (\vec r)$ with the local electric field $\vec E_{\rm
loc}(\vec r)$: \e
 \vec P(\vec r)=\=\alpha(\vec r)\vec E_{\rm loc}(\vec r).
\f

The expression for local polarizability in terms of dielectric
permittivity has the following form: \e \=\alpha^{-1}(\vec
r)=\left[\=\varepsilon(\vec r)-\varepsilon_0\=
I\right]^{-1}+\=I/(3\varepsilon_0), \l{alphar} \f where
$\varepsilon_0$ is permittivity of free space, $\=I$ is unit dyadic.
This expression follows from the Lorentz-Lorenz formula
\cite{BornWolf} \e \vec E_{\rm loc} (\vec r) = \vec E(\vec r) + \vec
P (\vec r)/(3\varepsilon_0)\f and material equation \e \vec D (\vec
r)=\varepsilon_0\vec E(\vec r)+\vec P(\vec r)=\=\varepsilon(\vec r)
\vec E(\vec r).\l{me}
 \f

\subsection{Dispersion equation}
Following local field approach one can write down dispersion
equation for the crystal under consideration in the next integral
form: \e \vec P (\vec r)=\=\alpha(\vec r) \int\limits_V\=G_3(\vec
r-\vec r',\vec q)\vec P (\vec r')d\vec r', \ \forall \ \vec r\in V,
\l{disp2} \f where $V=V(\vec a, \vec b, \vec c)$ is volume of the
elementary lattice cell, $\=G_3(\vec r, \vec q)$ is the lattice
dyadic Green's function: \e \=G_3(\vec r, \vec
q)=\sum\limits_{n,s,l}\=G(\vec r-\vec a n-\vec b s-\vec c
l)e^{-j(q_xan+q_ybs+q_zcl)}, \l{G3} \f which takes into account
cell-to-cell polarization distribution determined by wave vector
$\vec q=(q_x,q_y,q_z)^T$: \e \vec P(\vec r+\vec a n+\vec b s+\vec c
l)=\vec P(\vec r)e^{-j(q_xan+q_ybs+q_zcl)}, \l{shiftp2} \f $\=G(\vec
r)$ is dyadic Green's function of free space: \e \=G(\vec r) =
(k^2\=I+\nabla\nabla)\frac{e^{-jkr}}{4\pi\varepsilon_0 r},
\l{gfree}\f $n,s,l$ are integer indices, $k$ is wave number of free
space. The integral in \r{disp2} is singular if the point
corresponding to vector $r$ is located inside of some polarized
dielectric. It has to be evaluated in the meaning of principal value
by excluding small spherical region around the singular point and
tending the radius of this region to zero \cite{YaghjianDGF}.

The dispersion equation \r{disp2} relates distribution of
polarization density $\vec P(\vec r)$ and wave vector $\vec q$
corresponding to the eigenmodes of the electromagnetic crystal. If
the distribution of average electric field $\vec E(\vec r)$ of a
crystal eigenmode  is known then the polarization density $\vec
P(\vec r)$ can be found directly using material equation \r{me}: \e
\vec P(\vec r)=[\=\varepsilon(\vec r)-\varepsilon_0]\vec E(\vec
r).\f The reverse operation is possible only for space regions
filled by dielectric with $\=\varepsilon(\vec r)\ne \varepsilon_0$.
The distribution of electric field in free space regions, if
required, have to be calculated using the next integral
representation \e \vec E (\vec r)=\int\limits_V\=G_3(\vec r-\vec
r',\vec q)\vec P (\vec r')d\vec r'.\f

For our proof of generalized Ewald-Oseen extinction principle we
have to transform dispersion equation \r{disp2} into the form
corresponding to summation by layers in $x$-direction. The
expression for lattice dyadic Green's function $\=G_3(\vec r, \vec
q)$ \r{G3} can be rewritten using summation over planes in the form:
\e \=G_3(\vec r, \vec q)=\sum\limits_{n=-\infty}^{+\infty}\=G_2(\vec
r-\vec a n) e^{-jq_xan}, \l{G3layer} \f where $\=G_2(\vec r)$ is the
grid dyadic Green's function: \e \=G_2(\vec
r)=\sum\limits_{s,l}\=G(\vec r-\vec b s-\vec c
l)e^{-j(q_ybs+q_zcl)}. \f

Applying Poisson summation formula by both indices $s$ and $l$ one
can express the grid dyadic Green's function in terms of the spatial
Floquet harmonics. This expansion is also called as spectral
representation: \e \=G_2(\vec r) = \sum\limits_{s,l}
\=\gamma_{s,l}^{{\rm sign}(x-a)} e^{-j(\vec k^{{\rm
sign}(x)}_{s,l}\.\vec r)}, \l{genfloquet2}\f where $$
\=\gamma^{\pm}_{s,l}=\frac{j}{2bc\varepsilon_0 k_{s,l}} [\-
k^{\pm}_{s,l}\times[\- k^{\pm}_{s,l}\times \= I]], \ \vec
k^{\pm}_{s,l}=(\pm k^x_{s,l}, k^y_{s}, k^z_{l} )^T,$$
$$k^y_{s}=q_y+\frac{2\pi s}{b}, k^z_{l}=q_z+\frac{2\pi l}{c},
k^x_{s,l}=\sqrt{k^2-(k^y_{s})^2-(k^z_{l})^2}.$$ The square root in
the expression for $k_{s,l}$ should be chosen so that ${\rm
Im}(\sqrt{\.})<0$. The sign $\pm$ corresponds to half spaces $x>a$
and $x<a$ respectively.

Using \r{G3layer} the dispersion equation \r{disp2} can be rewritten
in the following form which will be used later on: \e \vec P (\vec
r)=\=\alpha(\vec r) \sum\limits_{n=-\infty}^{+\infty}\int\limits_V
\=G_2(\vec r-\vec r'-\vec a n) \vec P (\vec r') e^{-jq_xan} d\vec
r'. \l{displayer2} \f

\subsection{Semi-infinite crystal}
\begin{figure}[h]
\centering \epsfig{file=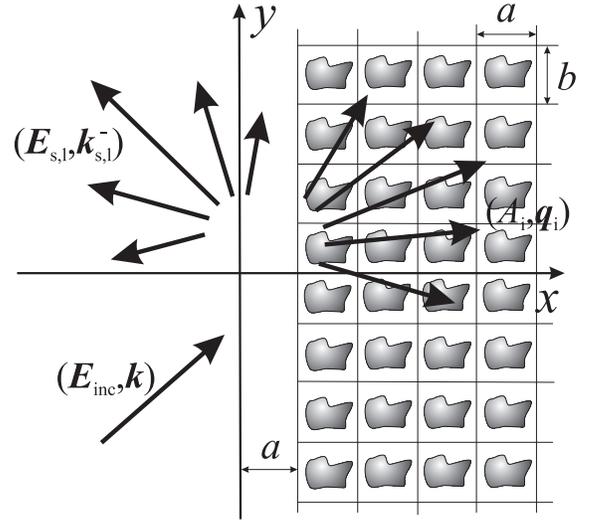, width=7.5cm} \caption{Geometry
of an semi-infinite electromagnetic crystal} \label{geomsemi}
\end{figure}

Now let us consider a semi-infinite crystal (half space $x\ge a$,
see Figure \ref{geomsemi}) excited by a plane electromagnetic wave
with wave vector $\vec k=(k_x,k_y,k_z)^{T}$ coming from free space:
\e \vec E_{\rm inc}(\vec r)=\vec E_{\rm inc} e^{-j(\vec k\.\vec r)}.
\l{inc2} \f The origin of our coordinate system is intentionally
shifted by one period into the free space since it simplifies rather
cumbersome calculations which are presented below and causes
exponential convergence of series in the final expressions.

Due to the periodicity of the semi-infinite structure along $y-$ and
$z-$axes the distribution of the excited polarization along these
directions is determined by the phase of the incident wave: \e \vec
P(\vec r+\vec a m+\vec b s+\vec c l)=\vec P(\vec r+\vec a
m)e^{-j(k_ybn+k_zcl)}, \l{shiftpsemi2} \f for any $\vec r\in V$ and
$m\ge 1$.

The electric field produced by the polarized semi-infinite crystal
has the form \e \vec E_{\rm scat}(\vec r) =
\sum\limits_{n=1}^{+\infty} \int\limits_V \=G_2(\vec r -\vec r' -
\vec a n)  \vec P(\vec r'+\vec a n) d\vec r'. \l{scat2} \f

The total local electric field is sum of incident and scattered
(produced by polarization of crystal) fields. Following the local
field approach we can write: \e \vec P(\vec r)= \=\alpha(\vec r)
\left[\vec E_{\rm inc}(\vec r)+\vec E_{\rm scat}(\vec r)\right].
\l{local} \f

Combining \r{scat2} and \r{local} we obtain an integral equation for
the polarization in the semi-infinite crystal excited by an incident
wave: \e \vec P(\vec r) =\=\alpha(\vec r) \left[ \vec E_{\rm
inc}(\vec r)+\sum\limits_{n=1}^{+\infty} \int\limits_V \=G_2(\vec r
-\vec r' - \vec a n)  \vec P(\vec r'+\vec a n) d\vec r'\right].
\l{layer2} \f

Now let us suppose that the dispersion equation \r{disp2} is solved
under condition that wave vector has the form $\vec
q=(q_x,k_y,k_z)^T$ with unknown $x-$component $q_x$, and the set of
eigenmodes $\{(\vec P_i(\vec r),q^{(i)}_x)\}$, characterized by
x-component of wave vector $q^{(i)}_x$ and polarization distribution
$\vec P_i(\vec r)$, is found. In addition, we include into this set
only the eigenmodes which either transfer energy into half-space
$x\ge a$ [$dq^{(i)}_x/d\omega>0$] or decay in $x$-direction [${\rm
Im}(q^{(i)}_x)<0$]. In such a case the polarization of the
semi-infinite crystal excited by the incident wave \r{inc2} can be
expanded by the eigenmodes of infinite crystal as follows: \e \vec
P(\vec r+\vec a m)=\sum\limits_i A_i \vec P_i(\vec r)
e^{-jq^{(i)}_xam}, \ \forall \vec r \in V, m\ge 1. \l{expand2} \f

Substituting \r{expand2} into \r{layer2} and using \r{inc2} we
obtain:

 \e \sum\limits_i A_i \vec P_i(\vec
r)e^{-jq_x^{(i)}am} =\=\alpha(\vec r) \left[
\vphantom{\sum\limits_{n=0}^{+\infty} \int\limits_V} \vec E_{\rm
inc}e^{-j \vec k(\vec r+\vec a m)}\right. \l{subst1} \f
$$
\left. +\sum\limits_{n=1}^{+\infty} \sum\limits_i A_i \int\limits_V
\=G_2(\vec r -\vec r' - \vec a (n-m)) \vec P_i(\vec r')
e^{-jq_x^{(i)}an} d\vec r'\right].
$$

Splitting series in the dispersion equation \r{displayer2} one can
derive the following auxiliary relation: $$\vec P_i (\vec
r)e^{-jq_x^{(i)}am}=
$$
\e =\=\alpha(\vec r) \sum\limits_{n=1}^{+\infty} \int\limits_V
\=G_2(\vec r-\vec r'-\vec a (n-m)) \vec P_i (\vec r')
e^{-jq_x^{(i)}an} d\vec r'  \l{aux2} \f
$$ +\=\alpha(\vec r) \sum\limits_{n=-\infty}^{0} \int\limits_V
\=G_2(\vec r -\vec r'-\vec a (n-m)) \vec P_i (\vec r')
e^{-jq_x^{(i)}an} d\vec r'.$$

Substituting \r{aux2} into \r{subst1} and following the fact that
${\rm det} \{\=\alpha(\vec r)\}\ne 0$ we obtain: $$ \sum\limits_i
A_i \sum\limits_{n=-\infty}^{0} \int\limits_V \=G_2(\vec r -\vec
r'-\vec a (n-m)) \vec P_i (\vec r') e^{-jq_x^{(i)}an} d\vec r' $$\e
=\vec E_{\rm inc}e^{-j \vec k(\vec r+\vec a m)}. \l{subst2} \f

Further, substituting \r{genfloquet2} into \r{subst2}, changing the
summation order and evaluating the sum of the geometrical
progression by index $n$ we get: $$ \sum\limits_{s,l} \left(
\sum\limits_i A_i \frac{\=\gamma_{s,l}^{+}  \int\limits_V \vec P_i
(\vec r') e^{j(\vec k^{+}_{s,l}\.\vec r')}d\vec
r'}{1-e^{j(q_x^{(i)}-k^x_{s,l})a}}\right) e^{-j(\vec
k^{+}_{s,l}\.(\vec r+\vec a m))}$$\e =\vec E_{\rm inc}e^{-j \vec
k(\vec r+\vec a m)}. \l{system2} \f

\subsection{Generalized Ewald-Oseen extinction principle}

The left part of equation \r{system2} represents an expansion of the
right part into a spatial spectrum of Floquet harmonics. The right
part represents an incident spectrum of Floquet harmonics containing
only the single incident plane wave \r{inc2} with $\vec k=\vec
k^+_{0,0}$. Equating coefficients in the left and right parts of
\r{system2} we obtain: \e \=\gamma_{s,l}^{+} \sum\limits_i A_i
\frac{\int\limits_V \vec P_i (\vec r') e^{j(\vec k^{+}_{s,l}\.\vec
r')}d\vec r'}{1-e^{j(q_x^{(i)}-k^x_{s,l})a}}
=\left\{\begin{array}{lcl}\vec E_{\rm inc},\ (s,l)=(0,0)\\ 0,\
(s,l)\ne (0,0) \end{array}.\right. \l{ewald2} \f

The values in the right side of \r{ewald2} are the amplitudes of the
incident spatial harmonics (all harmonics except fundamental one
have zero amplitudes), and the series in the left side are the
amplitudes of the spatial harmonics produced by the whole
semi-infinite crystal polarization in order to cancel these incident
harmonics. It means that equation \r{ewald2} represents the
generalization of Ewald-Oseen extinction principle (see \cite{Ewald,
Oseen, BornWolf} for classical formulation in the case of
dielectrics): {\it the polarization in a semi-infinite
electromagnetic crystal excited by a plane wave is distributed in
such a way that it cancels the incident wave together with all
high-order spatial harmonics associated with periodicity of the
boundary} (even if they have zero amplitudes as in the present
case). The additional words related to high-order Floquet harmonics
is the main and principal difference of Ewald-Oseen extinction
principle formulation for electromagnetic crystals as compared to
the classical case of isotropic dielectrics.

Substitution of \r{expand2} and \r{genfloquet2} into \r{scat2}
allows to express the scattered field in the half space $x<a$ in
terms of spatial Floquet harmonics: \e \vec E_{\rm scat}=
\sum\limits_{s,l} \vec E^{s,l}_{\rm scat} e^{-j(\vec k^-_{s,l}\.\vec
r)},\f  \e \vec E^{s,l}_{\rm scat}= \=\gamma_{s,l}^{-} \sum\limits_i
A_i \frac{\int\limits_V \vec P_i (\vec r') e^{j(\vec
k^{-}_{s,l}\.\vec r')}d\vec r'}{1-e^{j(q_x^{(i)}+k^x_{s,l})a}}.
\l{amplscat2} \f Note, that the formula for the amplitudes of
scattered Floquet harmonics \r{amplscat2} contains series which have
the same form as \r{ewald2} and differs only by the sign of the $x-$
components of wave vectors $\vec k^\pm_{s,l}=(\pm
k^x_{s,l},k_y,k_z)^T$ corresponding to the spatial harmonics
propagating into the half spaces $x<a$ and $x>a$, respectively.

If the eigenmodes of the crystal $\{q^{(i)}_x,\vec P_i(\vec r)\}$
are known then one can solve the system of linear equations
\r{ewald2} and find amplitudes of excited eigenmodes $\{A_i\}$. With
use of these amplitudes the scattered field can be found by
\r{amplscat2}. This provides a new numerical method which allows to
solve problem of plane-wave diffraction by a semi-infinite
electromagnetic crystal using knowledge of eigenmodes of the
infinite crystal. This fact is very important since at the moment
the reflection and dispersion problems for electromagnetic crystals
are usually solved by separate numerical approaches. The expressions
\r{ewald2} and \r{amplscat2} create a link between these two
problems and show how results of dispersion studies can be used in
order to describe reflection properties of electromagnetic crystals.

Until this point we considered only one incident wave with wave
vector $\vec k=\vec k^+_{0,0}$, but from \r{system2} it is clear
that we could consider also other incident spatial harmonics with
wave vectors $\vec k^+_{s,l}$ and get the similar results as
\r{ewald2}, but the nonzero terms at the right side of equation
would correspond to the respective incident spatial harmonic. Using
principle of superposition we obtain that if the semi-infinite
crystal is excited by whole spectrum of incident spatial harmonics
with amplitudes $\vec E^{s,l}_{\rm inc}$ and wave vectors $\vec
k^+_{s,l}$ then the following system of linear equations is valid:
\e \=\gamma_{s,l}^{+} \sum\limits_i A_i \frac{\int\limits_V \vec P_i
(\vec r') e^{j(\vec k^{+}_{s,l}\.\vec r')}d\vec
r'}{1-e^{j(q_x^{(i)}-k^x_{s,l})a}} =\vec E^{s,l}_{\rm inc}
\l{ewald3} \f

The scattered field is given by \r{amplscat2} as in the case of one
incident wave. The equation \r{ewald3} represents cancelation of all
incident spatial spectrum by induced polarization of the crystal in
accordance to the formulated above generalized Ewald-Oseen
extinction principle.

\subsection{Formulation of boundary conditions} In the previous
section we have proved generalized Ewald-Oseen extinction principle
for the case of semi-infinite electromagnetic crystal described by
certain periodic permittivity distribution $\=\varepsilon(\vec r)$
excited by a plane wave coming from free space. In order to extend
this theory for the case when incident wave comes from homogeneous
isotropic dielectric with permittivity $\epsilon$ it is enough to
change $\epsilon_0$ in all formulae to $\epsilon$. Physically it
means that we have to consider the polarization of the crystal with
respect to the host material with permittivity $\varepsilon$, but
not free space. In the model of dense cubic lattice of point dipoles
it means that the lattice is located inside of this host material.
This approach is very unusual since it can lead to results which are
strange from first point of view. For example, free space happen to
have negative polarization density with respect to dielectrics with
$\varepsilon>\varepsilon_0$. This can be simply explained since free
space with respect to these dielectrics is like real materials with
$\varepsilon<\varepsilon_0$ with respect to free space: they indeed
have negative polarization density.

The meaning of polarization density becomes relative automatically
when the replacement of $\varepsilon_0$ to $\varepsilon$ is made. In
order to avoid the use of this ambiguous polarization in the final
formulae it is possible to express the polarization density in terms
of the average field using \r{me}. The resulting expressions provide
complete set of boundary conditions for interface between
semi-infinite electromagnetic crystal and isotropic dielectric:

\e \vec E(\vec r)=\left\{ \begin{array}{lcl} \sum\limits_{s,l}
\left( \vec E^{s,l}_{\rm inc}e^{-j \vec k^-_{s,l}\vec r}+
\vec E^{s,l}_{\rm scat}e^{-j \vec k^+_{s,l}\vec r}\right),\ x<a\\
\sum\limits_{i} A_i \vec E_i e^{-j\vec q_i\vec r},\ x\ge a \\
\end{array} \right.
\f

\e \=\gamma_{s,l}^{+} \sum\limits_i A_i \frac{\int\limits_V
[\=\varepsilon-\varepsilon_0\=I]\vec E_i (\vec r') e^{j(\vec
k^{+}_{s,l}\.\vec r')}d\vec r'}{1-e^{j(q_x^{(i)}-k^x_{s,l})a}}=\vec
E_{\rm inc}^{s,l}, \l{bc1} \f

\e \=\gamma_{s,l}^{-} \sum\limits_i A_i \frac{\int\limits_V
[\=\varepsilon-\varepsilon_0\=I]\vec E_i (\vec r') e^{j(\vec
k^{-}_{s,l}\.\vec r')}d\vec r'}{1-e^{j(q_x^{(i)}+k^x_{s,l})a}}=\vec
E^{s,l}_{\rm scat}, \l{bc2} \f where we use following notations
$$ \=\gamma^{\pm}_{s,l}=\frac{j}{2bc\varepsilon k_{s,l}} [\-
k^{\pm}_{s,l}\times[\- k^{\pm}_{s,l}\times \= I]], \ \vec
k^{\pm}_{s,l}=(\pm k^x_{s,l}, k^y_{s}, k^z_{l} )^T,$$
$$k^y_{s}=q_y+\frac{2\pi s}{b}, k^z_{l}=q_z+\frac{2\pi l}{c},
k^x_{s,l}=\sqrt{k^2-(k^y_{s})^2-(k^z_{l})^2},$$ $k$ is wave number
in the dielectric with permittivity $\varepsilon$, $\vec
q_i=(q_i^x,k_y,k_z)^T$ and the square root in the expression for
$k_{s,l}$ should be chosen so that ${\rm Im}(\sqrt{\.})<0$.

The expressions \r{bc1} and \r{bc2} relate amplitudes of incident
$\vec E_{\rm inc}^{s,l}$ and scattered $\vec E_{\rm scat}^{s,l}$
spatial harmonics corresponding to tangential wave vector $\vec
k_t=(k_y,k_z)^T$ and periodicity of the boundary (rectangular
lattice with periods $b$ and $c$) with amplitudes $A_i$ of
eigenmodes $(\vec E_i,q^x_i)$ excited in the semi-infinite crystal.

The presented set of boundary conditions is complete: these
equations are enough to determine amplitudes of excited eigenmodes
and scattered spatial harmonics if the eigenmodes $\{(\vec
E_i,q_i^x)\}$ of infinite crystal corresponding to tangential wave
vector $\vec k_t$ are known. But this set is not unique. One can
immediately suggest to use classical boundary conditions (continuous
tangential component of electric field and normal component of
electric displacement at any point of the boundary $x=a$) which
being expanded into Fourier series will also grant a complete set of
linear equations relating amplitudes of incident, scattered and
excited modes.

The advantage of equations \r{bc1} and \r{bc2} as compared to any
other boundary conditions is such that they have very special form
which can admit analytical solution. We will demonstrate it in the
next section for the special case of electromagnetic crystals formed
by small scatterers which can be treated as point dipole with fixed
orientation. But this is not the only case when an analytical
solution of  \r{bc1} and \r{bc2} can be obtained. Recently, M.
Silveirinha \cite{MarioABC} demonstrated that the method proposed by
ourselves can be successfully applied for studies of reflection from
semi-infinite wire medium, material with strong low-frequency
spatial dispersion \cite{WMPRB}. Unfortunately, we can not provide
solution of equations \r{bc1} and \r{bc2} for the general case.
However, we can give some recommendations and an example how these
equations can be solved using method of characteristic function. We
hope that with some modification this method can be used for other
special cases as well.

\section{Lattice of uniaxial dipolar scatterers}

If the scatterers which form electromagnetic crystal are small as
compared to the wavelength then sometimes they can be effectively
replaced by point dipoles. It is assumed that the dipole moment of
such a dipole is determined by local field acting to the scatterer
and the field produced by the scatterer is equal to the field
created by the dipole. The polarizability which relates the induced
dipole moment and the local field acting to the scatterer is the
only parameter which depends on the shape of scatterer in such a
local field approach.

Generally, the field produced by any scatterer can be presented
using expansion by multipoles. The electric and magnetic dipoles are
first and second order multipoles. For some scatterers, the electric
or magnetic dipole moments dominate over high-order multipoles. It
means that some scatterers behave as electrical dipoles, some other
ones as magnetic. Below we will consider only such scatterers. The
scatterers which has both electric and magnetic dipole moments of
the same order or whose quadrupole or other high-order multipoles
can not be neglected are out of scope of our consideration.
Moreover, below we will consider only scatterers which can be
replaced by dipoles with fixed orientation.

\begin{figure}[h]
\centering \epsfig{file=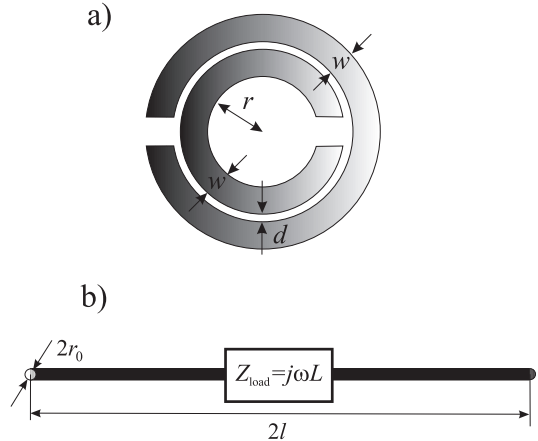, width=7cm} \caption{Geometries of
scatterers which can be modeled as dipoles with fixed orientation:
a) split-ring-resonator, b) inductively loaded wire.} \label{scat}
\end{figure}

The typical example of the scatterer which behaves as electric
dipole with fixed orientation at microwave frequencies is a short
metallic cylinder or piece of wire which can be loaded by some
inductance in order to increase its polarizability \cite{LWD} (see
Fig. \ref{scat}.b). At optical frequencies it can be prolate
metallic cylinder which has strong plasmonic resonance. The typical
magnetic scatterer at microwave frequencies is split-ring resonator
\cite{PendryMagnet} (see Fig. \ref{scat}.a) if the bianisotropic
properties of this scatterer are neglected or canceled using method
suggested in \cite{MarquesSRR}. At optical frequencies the split
metallic rings \cite{YenTHZ,LindenTHZ} behave as magnetic scatterers
with fixed orientation of dipole moment. Metallic spheres which also
can be replaced by point dipoles are out of scope of our
consideration since the orientation of their dipole moments depends
on direction of external field.

\subsection{History of the problem}

Further in this section we present an analytical solution for
problem of plane wave diffraction on semi-infinite electromagnetic
crystals formed by point scatterers with known polarizabilites, but
before that we have to describe history of this problem. The first
attempt to obtain such an analytical solution has been made by G.D.
Mahan and G. Obermair in a seminal work \cite{Mahan}. Analytical
expressions for reflection coefficients and amplitudes of excited
modes for a semi-infinite crystal were obtained in terms of wave
vectors of the infinite crystal eigenmodes. However this theory is
not free from the drawbacks. Mahan and Obermair treated the
interaction between a reference crystal plane of a semi-infinite
crystal and its $N$ nearest neighbors exactly, neglecting the other
crystal planes. That is why this approximation is called
``near-neighbor approximation''. Such an approach allows to
introduce fictitious zero polarization at the imaginary crystal
planes in free space over the semi-infinite crystal. This
manipulation gave a set of equations which were treated in
\cite{Mahan} as additional boundary conditions. It will be shown
below that if the interaction between planes is taken into account
exactly but not restricted to finite number of neighboring planes
then the fictitious polarization of imaginary planes turns out to be
non-zero. In works of C.A. Mead \cite{MeadExactly,MeadFormally} it
was already shown that the ``near-neighbor approximation'' appears
to be not a strict one. The mead states that the serious
disagreement appears in the cases then the interaction between
crystal planes falls off not sufficiently fast with distance. In
other words, the results of Mahan and Obermair are valid only then
the high-order spatial Floquet harmonics produced by the planes
rapidly decay with distance. Mahan and Obermair considered only the
normal incidence of the plane wave. Within such a restriction their
approach is valid in the case when the periods of the structure are
small as compared with the wavelength in the host medium. The strong
disagreement with the exact solution appears when one of high-order
Floquet harmonic happens to be propagating one. This fact is
illustrated below by a numerical comparison.

The work \cite{Mahan} caused numerous extensions
\cite{PhilpottRef,PhilpottSlab,PhilpottPRB}. The Mahan and Obermair
approach was generalized for the cases of oblique incidence
\cite{PhilpottRef}, both possible polarizations of incident wave
\cite{PhilpottPRB}, various lattice structures of the crystal
\cite{PhilpottRef}, tensorial polarizability of scatterers
\cite{PhilpottPRB} and even diffraction of the finite-size slabs of
the crystals was considered \cite{PhilpottSlab}. Note, that all the
listed works use the same ``near-neighbor approximation'' and their
applicability is restricted as described above. In order to avoid
this trouble one needs to use another model for interaction between
crystal planes. The simplest one is the so-called ``exp model''
suggested by Mead \cite{MeadExactly,MeadExp2} which assumes that
interaction can be described by a single decaying exponent. In terms
of spatial Floquet harmonics this approach is equivalent to
neglecting all high-order Floquet harmonics except the one with the
slowest decay. The ``exp model'' as well as the ``near-neighbor
approximation'' allow to solve the problem of excitation
analytically for both normal \cite{MeadExactly} and oblique
\cite{MeadExp2} incidences. The ``exp model'' of Mead gives a set of
two equations which correspond to the generalized Ewald-Oseen
extinction principle formulated in the present paper. The first
equation of Mead is the same as one of equations given by
``near-neighbor approximation''. It describes the fact that the
incident electromagnetic wave (fundamental Floquet harmonic) inside
the semi-infinite crystal is canceled by induced polarization of the
crystal. This fact was pointed out in the papers
\cite{Mahan,PhilpottPRB,MeadExactly}. The second equation clearly
expresses the fact that induced polarization cancels also the second
Floquet harmonic (taken into account in the ``exp model'') of
incident wave which has zero amplitude, but unfortunately it was not
noted by the authors. The system of these two equations is solved in
\cite{MeadExactly} and the amplitudes of excited eigenmodes and an
expression for reflection coefficient are obtained.

It is possible to modify the ``exp model'' in order to obtain an
exact solution. For that purpose one simply should take into account
all Floquet harmonics in the interaction between the crystal planes.
This has been done by authors of the present paper and the results
are presented below. As it was shown above, it turns out that every
incident Floquet harmonic (even if it has zero amplitude) is
canceled by the induced polarization following to the generalized
Ewald-Oseen extinction principle. It provides an infinite system of
equations relating amplitudes of excited eigenmodes. This system can
be truncated and then the number of equations in the system turns
out to be equal to the number of Floquet harmonics taken into
account. Such the finite system can be easily solved analytically
for the case when only two Floquet harmonics are taken into account
(this is the ``exp model'' of Mead \cite{MeadExactly}), but in the
case when one would like to take into account more Floquet harmonics
this approach requires numerical calculations. We avoid the
truncation of the system of equations and offer a closed-form
rigorous analytical solution which is simple and explicit.

Note, that a ``formally closed solution'' for the problem under
consideration was proposed by Mead in \cite{MeadFormally}. In this
solution there is a contour integral of a certain function given in
the form of infinite series. However, the calculation with help of
such ``formally closed solution'' requires serious numerical
efforts. The main idea of work \cite{MeadFormally} is based on
introduction of characteristic analytical function  which allows to
determine all parameters entering the expression for the reflection
coefficient. It is shown that knowledge of its roots allows to
recover this function and obtain analytical expressions for all
amplitudes of excited eigenmodes and for the reflection coefficient,
consequently. Unfortunately, these roots were not found in
\cite{MeadFormally}. That is why the contour integration was used in
\cite{MeadFormally} in order to bypass the problem of these roots
finding. In fact, as it is shown below, the roots of this
characteristic analytical function are determined by the wave
vectors of Floquet harmonics and can be easily expressed
analytically. This fact is a consequence of generalized Ewald-Oseen
extinction principle which is an important point of our theory.

One could directly apply general results \r{ewald2} and
\r{amplscat2} to the problem under consideration. It is enough to
replace polarization density of eigenmodes by three-dimensional
delta function corresponding to the point of location of the dipolar
scatterer in the unit cell and one could obtain the system of linear
equations \r{ewald2} and \r{amplscat2} which correspond to the
boundary conditions in the present case as it is shown below.
However, we prefer to re-deduce all the expressions by making the
same steps as in the previous section but for the case of point
scatterers. We suppose that it is very useful step in order to
demonstrate physical background of the computations undertaken in
the previous section at a specific example.

\subsection{Dispersion equation}

\begin{figure}[h]
\centering \epsfig{file=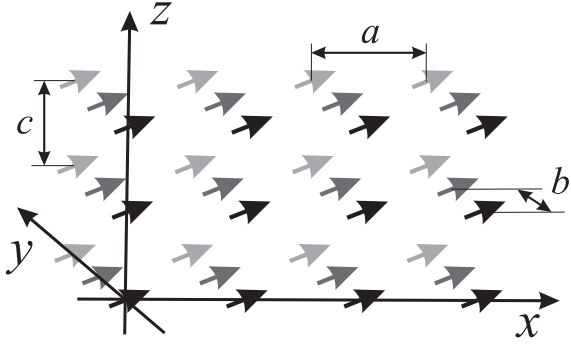, width=7.5cm} \caption{Geometry of
an infinite electromagnetic crystal formed by uniaxial dipolar
scatterers} \label{geomsm}
\end{figure}

Let us consider an infinite crystal formed by point dielectric
dipoles with some known polarizability $\alpha$ along fixed
direction given by unit vector $\vec d$, $\=\alpha=\alpha \vec d\vec
d$. The case of magnetic dipoles can be easily obtained from the
theory for dielectric ones using duality principle. The scatterers
are arranged in the nodes of the three-dimensional lattice with an
orthorhombic elementary cell $a\times b\times c$ located in free
space, see Figure \ref{geomsm}.

The distribution of dipole moments corresponding to an eigenmode
with wave vector $\vec q=(q_x,q_y,q_z)^T$ is described as $\vec
p_{n,s,l}=\vec p e^{-j(q_xan+q_ybs+q_zcl)}$, where $n,s,l$ are
integer indices of scatterers along the $x-,y-,z-$ axes,
respectively, and $\vec p$ is a dipole moment of the scatterer
located at the center of coordinate system. Following to the local
field approach $\vec p$ can be expressed as $\vec p=\alpha (\vec
E_{\rm loc}\.\- d)\- d$, where $\vec E_{\rm loc}$ is a local
electric field acting to the scatterer. The local field is produced
by all other scatterers which form the infinite crystal and it can
be given by the formula \e \vec E_{\rm
loc.}={\sum\limits_{n,s,l}}'\=G(\vec R_{n,s,l})\vec p_{n,s,l}, \f
where $\=G(\vec r)$ is the three-dimensional dyadic Green function
of the free space \r{gfree} and summation is taken over all triples
of indices except the zero one. Accordingly the following dispersion
equation for the crystal under consideration is obtained (compare
with \r{disp2}): \e \alpha^{-1}=\left[{\sum\limits_{n,s,l}}'\=G(\vec
R_{n,s,l})e^{-j(q_xan+q_ybs+q_zcl)}\-d\right]\.\-d. \l{gendis} \f

In order to evaluate sums of series in \r{gendis} we use a
plane-wise approach \cite{nonresPRE}. According to this approach the
dispersion equation takes the following form: \e
\alpha^{-1}=\sum\limits_{n=-\infty}^{+\infty} \beta_n e^{-jq_xan}.
\l{layerdisp} \f

The coefficients $\beta_n$ describe the interaction between planes
and include the information on transverse wave vector components
$q_y$, $q_z$ as well as on a geometry of a single plane (lattice
periods $b$, $c$). For $n\ne 0$ coefficients $\beta_n$ can be
expressed using expansion by Floquet harmonics. For $n=0$ the
calculation of coefficient $\beta_0$ (describing interaction inside
of a plane and expressed in the form of two-dimensional series
without the zero term) requires additional efforts (see
\cite{nonresPRE,Belovhomo,Collin} for details).

The electric field produced by a single plane (namely
two-dimensional grid $b\times c$ of point dipoles with the
distribution $\vec p_{s,l}=\vec p e^{-j(q_ybs+q_zcl)}$) located in
the plane $x=0$ is equal to \e \vec E(\vec r) =
\frac{j}{2bc\varepsilon_0}\sum\limits_{s,l} [\- k^{{\rm
sign}(x)}_{s,l}\times[\- k^{{\rm sign}(x)}_{s,l}\times \vec p]]
\frac{e^{-j(\vec k^{{\rm sign}(x)}_{s,l}\.\vec R)}}{k^x_{s,l}},
\l{genfloquet} \f where $\vec k^{\pm}_{s,l}=(\pm k^x_{s,l}, k^y_{s},
k^z_{l} )^T$, $k^y_{s}=q_y+\frac{2\pi s}{b}$,
$k^z_{l}=q_z+\frac{2\pi l}{c}$,
$k^x_{s,l}=\sqrt{k^2-(k^y_{s})^2-(k^z_{l})^2}$ and$k$ is wave number
of free space. One should choose the square root in the expression
for $k_{s,l}$ so that ${\rm Im}(\sqrt{\.})<0$. The sign $\pm$
corresponds to half spaces $x>0$ and $x<0$ respectively.

The formula \r{genfloquet} defines an expansion of the field
produced by a single grid of dipoles in terms of plane waves and it
can be obtained using double Poisson summation formula to series of
fields produced by single scatterers in free space. These plane
waves have wave vectors $\vec k^{\pm}_{s,l}$. They are also called
Floquet harmonics and represent a spatial spectrum of the field
(compare with \r{genfloquet2}). Floquet harmonics are widely used in
analysis of phased array antennas \cite{Amitay}.

Using \r{genfloquet} we get the following expression for $\beta_n$
($n\ne 0$): \e \beta_n=\sum\limits_{s,l}\gamma^{-{\rm
sign}(n)}_{s,l} e^{-jk^x_{s,l}a|n|}, \l{floquet} \f where
$\gamma^\pm_{s,l}=[k^2-(\- k^\pm_{s,l}\.\-d)^2]/(2jbc\varepsilon_0
k^x_{s,l})$. After substitution of \r{floquet} into \r{layerdisp},
changing the order of summation and using the formula for sum of
geometrical progression we obtain the dispersion equation in the
following form: \e \alpha^{-1}=\beta_0 + \sum\limits_{s,l} \left[
 \frac{\gamma^{-}_{s,l}}{e^{j(k^x_{s,l}+q_x)a}-1}+
 \frac{\gamma^{+}_{s,l}}{e^{j(k^x_{s,l}-q_x)a}-1}\right].
\l{disp} \f This is a transcendental equation expressed by rapidly
convergent series. Dispersion properties of the crystal under
consideration can be studied with help of numerical solution of the
latter equation. The case of dipoles oriented along one of the
crystal axes of the orthorhombic crystal has been considered in
\cite{Belovhomo}, the dispersion equation was solved and typical
dispersion curves and iso-frequency contours for resonant scatterers
were presented.

\subsection{Semi-infinite crystal}
\begin{figure}[ht]
\centering \epsfig{file=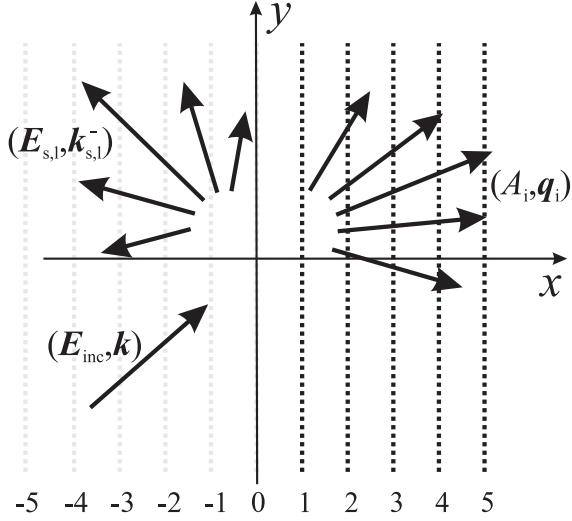, width=7.5cm}
\caption{Geometry of a semi-infinite electromagnetic crystal formed
by uniaxial dipolar scatterers} \label{semigeom}
\end{figure}

Now let us consider a semi-infinite electromagnetic crystal, the
half-space $x\ge a$ filled by the crystal formed by point dipoles,
see Figure \ref{semigeom}.
The structure is excited by a plane electromagnetic wave with the
wave vector $\-k=(k_x,k_y,k_z)^T$ and the intensity of electric
field $\- E_{\rm inc}$. Let us denote the component of the incident
electric field along the direction of dipoles as $E_{\rm inc}=(\-
E_{\rm inc}\.\-d)$. The axis $x$ is assumed to be normal to the
interface. The tangential (with respect to the interface)
distribution of dipole moments in excited semi-infinite crystal is
determined by the tangential component of the incident wave vector.
It means that $p_{n,s,l}=p_{n}e^{-j(k_ybs+k_zcl)}$, where the
polarizations of zero-numbered scatterers from planes with the index
$n$ (parallel to the interface) are denoted as $p_n=p_{n,0,0}$. The
plane-to-plane distribution $\{p_n\}$ is unknown and it has to be
found. Using the local field approach one can write the infinite
linear system of equations for this distribution: \e p_m =\alpha
\left(E_{\rm inc}e^{-jk_xam}+\sum\limits_{n=1}^{+\infty} \beta_{n-m}
p_n \right), \ \forall\ m\ge 1 . \l{layer} \f The distribution
$\{p_n\}$ of polarization in the excited semi-infinite crystal can
be determined solving the system of equations \r{layer}. The known
distribution of polarization allows to determine the scattered field
in the half-space $x<a$ with help of the expansion by Floquet
harmonics \r{genfloquet}: \e \vec E = \sum\limits_{s,l} \vec E_{s,l}
e^{-j(\vec k^-_{s,l}\.\vec R)}, \l{diffr} \f where the amplitudes of
Floquet harmonics are following: \e \vec E_{s,l} =
\frac{j}{2ab\varepsilon_0 k^x_{s,l}}[\- k^-_{s,l}\times[\-
k^-_{s,l}\times \vec d]] \sum\limits_{n=1}^{+\infty} p_n
e^{-jk^x_{s,l}an}. \l{ampl} \f

If the crystal supports propagating modes, it is quite difficult to
find a solution of \r{layer} numerically. Simple methods such as
system truncating (considering a slab with finite thickness instead
of half-space like in \cite{flo}) results in nonconvergent
oscillating solutions which have nothing to do with actual solution
of \r{layer}.

\subsection{Expansion by eigenmodes}
In order to solve \r{layer} accurately one has to use an expansion
of the polarization by eigenmodes \cite{Mahan}: \e p_n
=\sum\limits_{i} A_i e^{-jq_x^{(i)}an}, \l{expansion} \f where $A_i$
are amplitudes of eigenmodes and $q_x^{(i)}$ are the $x$-components
of their wave vectors. Every eigenmode is assumed to be a solution
of the dispersion equation \r{disp} with the wave vector $\vec
q_i=(q_x^{(i)},k_y,k_z)^{T}$. In the formula \r{expansion} the
summation is taken by eigenmodes which either transfer energy into
half-space $x\ge a$ ($\frac{dq_x^{(i)}}{d\omega}>0$) or decay along
$x-$axis (${\rm Im}(q_x^{(i)})<0$).

Let us assume that the dispersion equation \r{disp} is solved (for
example numerically) and the necessary set of eigenmodes
$\{q_x^{(i)}\}$ is found. Then the substitution of \r{expansion}
into \r{layer} will replace the set of unknown  polarizations of
planes by a set of unknown amplitudes of eigenmodes: \e
\alpha^{-1}\sum\limits_{i} A_i e^{-jq_x^{(i)}am} = E_{\rm
inc}e^{-jk_xam}+\sum\limits_{n=1}^{+\infty} \beta_{n-m}
\sum\limits_{i} A_i e^{-jq_x^{(i)}an}. \l{st} \f Applying the
auxiliary relation evidently following from \r{disp}: \e \alpha^{-1}
e^{-jq_x^{(i)}am}-\sum\limits_{n=-\infty}^{0} \beta_{n-m}
e^{-jq_x^{(i)}an}= \sum\limits_{n=1}^{+\infty} \beta_{n-m}
e^{-jq_x^{(i)}an}, \l{aux} \f the equation \r{st} can be transformed
as follows: \e \sum\limits_{i} A_i \left(\sum\limits_{n=-\infty}^{0}
\beta_{n-m} e^{-jq_x^{(i)}an} \right)=E_{\rm inc}e^{-jk_xam}.
\l{eigenamp} \f

It should be noted, that using definition of Mahan and Obermair for
the polarization of fictitious planes ($p_n=\sum\limits_{i} A_i
e^{-jq_x^{(i)}an}, \ \forall n\le 0$) one can rewrite \r{eigenamp}
as \e \sum\limits_{n=-\infty}^{0} \beta_{n-m} p_n=E_{\rm
inc}e^{-jk_xam}. \l{fictitious} \f It is evident that the assumption
of Mahan and Obermair, requiring all polarizations of fictitious
planes to be zeros, contradicts with \r{fictitious}. This fact
proves that the ``near-neighbor approximation'' made in \cite{Mahan}
is not accurate.

The system of equations \r{eigenamp} can be truncated and solved
numerically quite easily in contrast to \r{layer}. As a result, the
amplitudes of eigenmodes $\{A_i\}$ can be found and the polarization
distribution can be restored using formula \r{expansion}. The
amplitudes of scattered Floquet harmonics \r{ampl} can be also
expressed in terms of excited eigenmodes amplitudes by means of
substitution of \r{expansion} into \r{ampl}, changing the order of
summation and evaluating sums of geometrical progressions. The final
expression for the amplitudes of scattered Floquet harmonics is
following (compare with \r{amplscat2}) : \e \vec E_{s,l} = \frac{[\-
k^-_{s,l}\times[\- k^-_{s,l}\times \vec d]]}{2jab\varepsilon_0
k^x_{s,l}} \sum\limits_{i}
A_i\frac{1}{1-e^{j(q_x^{(i)}+k^x_{s,l})a}}. \l{ampl2} \f

One can stop at this stage and claim that the problem of the
semi-infinite electromagnetic crystal excitation is solved. However
in this case the solution would require long numerical calculations,
such as solving the system \r{eigenamp} and substituting the
obtained solution into \r{ampl2}. The possibility to make all
described operations analytically in the closed-form is shown below.

\subsection{Analytical solution}
Substituting the expansion \r{floquet} into \r{eigenamp} we obtain:
$$\sum\limits_{i} A_i \left(\sum\limits_{n=-\infty}^{0}
\left[\sum\limits_{s,l}\gamma^+_{s,l} e^{-jk^x_{s,l}a|n-m|}\right]
e^{-jq_x^{(i)}an} \right)$$ \e =E_{\rm inc}e^{-jk_xam}. \l{big} \f
Changing the order of summation in \r{big}, taking into account that
$n-m<0$ and using formula for the sum of geometrical progression we
obtain: \e \sum\limits_{s,l}\gamma^+_{s,l}  \left(\sum\limits_{i}
A_i \frac{1}{1-e^{j(q_x^{(i)}-k^x_{s,l})a}}\right)
e^{-jk^x_{s,l}am}=E_{\rm inc}e^{-jk_xam} \l{syst} \f This is a
system of linear equations where unknowns are given by expressions
in brackets. It has a unique solution because the determinant of the
system has finite nonzero value. Note that $k^x_{s,l}=k_x$ only if
$(s,l)=(0,0)$. Thus, the solution of \r{syst} has the following
form: \e \sum\limits_{i} A_i
\frac{1}{1-e^{j(q_x^{(i)}-k^x_{s,l})a}}=\left\{\begin{array}{lcl}
E_{\rm inc}/\gamma^+_{0,0},\ \mbox{if}\  (s,l)=(0,0)\\ 0, \
\mbox{if}\ (s,l)\ne (0,0)\\ \end{array}\right. \l{ewald} \f

This equation could be directly obtained from general expression
\r{ewald2} by substitution of delta function instead of polarization
density of eigenmodes, but as we already mentioned above we
intentionally re-deduced it since we suppose that it can help to
understand background for deduction of \r{ewald2} for general case.
The values at the right side of \r{ewald} are the normalized
amplitudes of incident Floquet harmonics, and the series at the left
side are the normalized amplitudes of Floquet harmonics produced by
the whole semi-infinite crystal polarization which cancel the
incident harmonics. Thus, the equation \r{ewald} represents the
generalization of the  Ewald-Oseen extinction principle already
formulated above for general case of semi-infinite crystals: {\it
The polarization in a semi-infinite electromagnetic crystal excited
by a plane wave is distributed in such a way that it cancels the
incident wave together with all high-order spatial harmonics
associated with periodicity of the boundary}.

Note, that the formula for the amplitudes of scattered Floquet
harmonics \r{ampl2} contains series that have the same form as
\r{ewald}, but another sign in front of $k^x_{s,l}$.

The amplitudes of the excited modes $A_i$ can be found numerically
from the infinite set of equations \r{ewald} and substitution of
$A_i$ into \r{ampl2} will give us amplitudes of scattered Floquet
harmonics. However, it is possible to obtain a closed-form
analytical solution of \r{ewald}.

In order to solve the set of equations \r{ewald} one should consider
a characteristic function $f(x)$ (see also \cite{MeadFormally}) of
the form : \e f(u)=u\sum\limits_{i} A_i \frac{1}{u-e^{jq_x^{(i)}a}}.
\l{f} \f Comparing \r{ampl2} and \r{ewald} with \r{f} one can see
that the function $f(u)$ has the following properties:
\begin{itemize}
\item It has poles at $u=e^{jq_x^{(i)}a}$
\item It has roots at $u=e^{jk^x_{s,l}a}$, $(s,l)\ne (0,0)$ and $u=0$
\item It has a known value $E_{\rm inc}/\gamma^+_{0,0}$ at $u=e^{jk_xa}$
\item Its values at $u=e^{-jk^x_{s,l} a}$ are equal to the normalized amplitudes of scattered Floquet harmonics
\item Its residues at $u=e^{jq_x^{(i)}a}$ are equal to the normalized amplitudes of excited eigenmodes.
\end{itemize}

It is possible to restore the function $f(u)$ using the known values
of its poles, roots and a value at one point: \e f(u)=\frac{E_{\rm
inc}u}{\gamma^+_{0,0} e^{jk_xa}} \prod\limits_{(s,l)\ne (0,0)}
\frac{u-e^{jk^x_{s,l}a}}{e^{jk_xa}-e^{jk^x_{s,l}a}} \prod\limits_i
\frac{e^{jk_xa}-e^{jq_x^{(i)}a}}{u-e^{jq_x^{(i)}a}}. \f

The knowledge of the characteristic function $f(u)$ provide us with
complete solution of our diffraction problem. The amplitudes of
excited eigenmodes with indices $n$ are equal to residues of $f(u)$
at $u=e^{jq_x^{(n)}a}$:
$$
A_n={\rm Res} f(u) \left|_{u=e^{jq_x^{(i)}a}} \vphantom{\frac{a}{a}}
\right.=\frac{E_{\rm
inc}(1-e^{j(q_x^{(n)}-k_x)a})}{\gamma^+_{0,0}}\times
$$
\e \times \prod\limits_{(s,l)\ne (0,0)}
\frac{e^{jq_x^{(n)}a}-e^{jk^x_{s,l}a}}{e^{jk_xa}-e^{jk^x_{s,l}a}}
\prod\limits_{i\ne n}
\frac{e^{jk_xa}-e^{jq_x^{(i)}a}}{e^{jq_x^{(n)}a}-e^{jq_x^{(i)}a}},
\l{A} \f and the amplitudes of scattered Floquet harmonics with
indices $(r,t)$ can be expressed through values of $f(u)$ at
$u=e^{-jk^x_{r,t} a}$:
$$
\-E_{r,t}= \frac{E_{\rm inc}e^{-jk^x_{r,t}a}[\- k^-_{r,t}\times[\-
k^-_{r,t}\times \vec d]]}{2jab\varepsilon_0 k^x_{r,t}\gamma^+_{0,0}
e^{jk_xa}} \times
$$
\e \times \prod\limits_{(s,l)\ne (0,0)}
\frac{e^{-jk^x_{r,t}a}-e^{jk^x_{s,l}a}}{e^{jk_xa}-e^{jk^x_{s,l}a}}
\prod\limits_i
\frac{e^{jk_xa}-e^{jq_x^{(i)}a}}{e^{-jk^x_{r,t}a}-e^{jq_x^{(i)}a}}.
\l{R} \f

The products in the formulae \r{A} and \r{R} have very rapid
convergence. It is enough to take a few terms in order to reach
excellent accuracy. The main requirement for truncation of these
infinite products is to take into account all terms corresponding to
propagating $\mbox{Re}(k^x_{s,l})=0$ and slowly decaying
$\mbox{Im}(k^x_{s,l})\ll 2\pi/a$ Floquet harmonics as well as
propagating $\mbox{Re}(q_x^{(i)})=0$ and slowly decaying
$\mbox{Im}(q_x^{(i)})\ll 2\pi/a$ eigenmodes.

\subsection{Comparison with other theories}

Let us consider the case from work \cite{Mahan} when $\-d=\-y_0$,
and $a=b=c$. In this case the formula \r{R} for the fundamental
Floquet harmonic ($r=t=0$) can be rewritten in terms of the
reflection coefficient: \e R= -e^{-2jk_xa} \prod\limits_{(s,l)\ne
(0,0)} \frac{e^{-jk_xa}-e^{jk^x_{s,l}a}}{e^{jk_xa}-e^{jk^x_{s,l}a}}
\prod\limits_i
\frac{e^{jk_xa}-e^{jq_x^{(i)}a}}{e^{-jk_xa}-e^{jq_x^{(i)}a}}.
\l{Rmain} \f

Comparing that result with the final result of the work \cite{Mahan}
(the next formula after (C7) on page 841) one can see that the first
product in our formula \r{Rmain} \e \Pi=\prod\limits_{(s,l)\ne
(0,0)} \frac{e^{-jk_xa}-e^{jk^x_{s,l}a}}{e^{jk_xa}-e^{jk^x_{s,l}a}}
\l{Pi} \f is absent  in the result of Mahan and Obermair. This
difference is a consequence of the fact that in our study we
considered interaction between crystal planes accurately taking into
account all Floquet harmonics for any distance between planes in
contrast to the ``near-neighbor approximation'' used in the approach
of Mahan and Obermair.

\begin{figure}[ht]
\vspace{5mm}
\centering
\epsfig{file=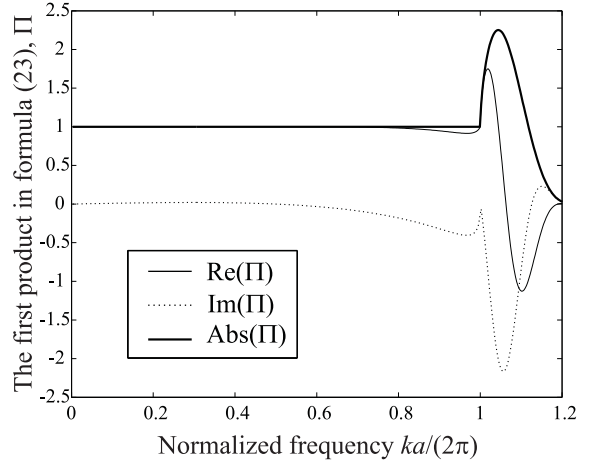, width=7.5cm}
\caption{Dependence of $\Pi$ vs. normalized frequency $ka/(2\pi)$}
\label{prod}
\end{figure}

The dependence of the product $\Pi$ vs. normalized frequency is
plotted in Figure \ref{prod} for the case of normal incidence
$k_y=k_z=0$ and $k_x=k$. One can see that the value of the product
is nearly equal to the unity for $ka<1.6\pi$, but for $ka>2\pi$ the
value of the product significantly differs from the unity. Thus we
conclude that the theory of Mahan and Obermair is valid in the low
frequency range when periods of the lattice are small compared to
the wavelength. Our theory does not have such a restriction (within
the frame of the dipole model of electromagnetic crystal).

The comparison with results of \cite{MeadFormally} shows that
\r{Rmain} is equivalent to formula (46) from \cite{MeadFormally}
with $\Pi=\exp(\Gamma)$ where $\Gamma$ is given by the contour
integral (47) from \cite{MeadFormally}. The calculation of $\Pi$
using \r{Pi} requires taking into account only a few terms in the
infinite products, because they are very rapidly convergent. This is
a significant advantage of our approach as compared to work
\cite{MeadFormally} which requires complicate numerical calculation
of the contour integral.

In the long-wavelength limit, the series in equation \r{gendis} for
the cubic lattice can be replaced by the integral taken over the
whole space except unit cell $V$ and we obtain: \e
\alpha^{-1}=\left[\left(\int\limits_{R^3\slash V}\=G(\vec
R)e^{-j(\vec q\. \vec R)} d\vec R\right)\-d\right]\.\-d. \l{disint}
\f

The integral in the right-hand side of equation \r{disint} can be
evaluated by means of the same technique that was used while
deducing Ewald-Oseen extinction principle in \cite{BornWolf}. The
result is following: \e \alpha^{-1}=\left[\frac{1}{3}+\frac{(\vec
q\. \vec d)^2-|\vec q|^2}{K^2-|\vec
q|^2}\right]\frac{V}{\varepsilon_0}. \l{disint2} \f

The obtained dispersion equation \r{disint2} can be transformed in
the common form: \e \tilde\varepsilon (k^2-q_d^2)=\varepsilon_0
(|\vec q|^2-q_d^2), \l{disuni} \f where \e
\tilde\varepsilon=\varepsilon_0\left(1+\frac{\alpha/(\varepsilon_0
V)}{1-\alpha/(3\varepsilon_0 V)}\right), \l{epseff} \f and
$q_d=(\vec q\. \vec d)$ is the component of the wave vector $\vec q$
along the anisotropy axis.

The formula \r{disuni} is classical form of the dispersion equation
for uniaxial dielectrics \cite{BornWolf} with permittivity
$\tilde\varepsilon$ along the anisotropy axis and $\varepsilon$ in
the transverse plane. The expression \r{epseff} is the
Clausius-Mossotti formula for the effective permittivity of cubic
lattices of scatterers.

In the long-wavelength limit the formula \r{R} for amplitude of
reflected wave simplifies as follows: \e \-E_R= -\frac{(\vec E_{\rm
inc}\.\vec d) [\- k^-\times[\- k^-\times \vec d]]}{[k^2-(\-
k\.\-d)^2]} \frac{k_x-q_x}{k_x+q_x}, \l{Rlong} \f where $\vec
k^-=(-k_x,k_y,k_z)^T$ is wave vector of reflected wave. The formula
\r{Rlong} represents a compact form of an expression for electric
field amplitude of a wave reflected from an interface between an
isotropic dielectric and an uniaxial dielectric (see, e.g.
\cite{Fedorov}). Note, that in our case the situation is simplified
as compared to the general case, because the incident wave comes
from isotropic dielectric with permittivity $\varepsilon$ which is
equal to the permittivity of uniaxial dielectric in transverse
plane. It means, that an incident wave with normal polarization with
respect to the anisotropy axis transforms at the interface in a
refracted ordinary wave without reflection.

Let us consider the reflection problem at the special case when
$\-d=\-y_0$, $k_y=0$ and $\vec E_{\rm inc}\parallel \-y_0$. The
nonzero components of the wave vector for the incident wave can be
expressed in terms of incident angle $\theta_i$ as
$k_z=k\sin\theta_i$ and $k_x=k\cos\theta_i$. From \r{disuni} we
obtain that in this case the transmitted wave has $x$-component of
wave vector equal to $q_x=\sqrt{\tilde\varepsilon
k^2/\varepsilon_0-k_z^2}=\sqrt{\tilde\varepsilon/\varepsilon_0}k\cos\theta_t$,
where $\theta_t$ in angle or refraction. With the result \r{Rlong}
we get the reflection coefficient in the form: \e
R=\frac{k_x-q_x}{k_x+q_x}=\frac{n_1\cos\theta_i-n_2\cos\theta_t}{n_1\cos\theta_i+n_2\cos\theta_t},
\l{Fresnel} \f where $n_1=\sqrt{\varepsilon_0\mu_0}$ and
$n_2=\sqrt{\tilde\varepsilon\mu_0}$ are indices of refraction of the
materials. The formula \r{Fresnel} coincide with the classical
Fresnel equation \cite{BornWolf}. This fact can be treated as an
additional verification of presented theory.

\section{Lattice of split-ring-resonators}

In this section we apply theory presented in previous sections for
study of reflection from semi-infinite cubic lattice of
split-ring-resonators.

The general dispersion equation \r{disp2} of the integral form in
the case of point electric scatterers transforms into transcendental
equation \r{disp}. In the case when $\vec d=\vec y_0$ the dispersion
equation \r{disp} can be rewritten in the following closed form
convenient for numerical calculations (see \cite{Belovhomo} for
details): \e \varepsilon_0\alpha^{-1}(\omega)=C(k,\-q,a,b,c),
\l{disphomo}\f where $C(k,\-q,a,b,c)$ is dynamic interaction
constant of the form:
$$
C(k,\-q,a,b,c)= -\sum\limits_{l=1}^{+\infty}\sum\limits_{{\rm
Re}(p_s)\ne 0} \frac{p_s^2}{\pi b}K_0\left(p_scl\right)\cos(q_zcn)
$$
\e
+\sum\limits_{s=-\infty}^{+\infty}\sum\limits_{l=-\infty}^{+\infty}
\frac{p_s^2}{2jbc k^x_{s,l}} \frac{e^{-j k^x_{s,l} a}-\cos
q_xa}{\cos k^x_{s,l} a-\cos q_xa} \l{cfinal} \f
$$
-\sum\limits_{{\rm Re}(p_s)=0} \frac{p_s^2}{2bc}
\left(\frac{1}{jk^x_{s,0}}+ \sum\limits_{l=1}^{+\infty}
\left[\frac{1}{jk^x_{s,l}}+\frac{1}{jk^x_{s,-l}}\right. \right.
$$
$$
\left.\left. -\frac{c}{\pi l}-\frac{r_sc^3}{8\pi^3 l^3}\right]+1.202
\frac{r_sc^3}{8\pi^3}+ \frac{c}{\pi} \left(\log
\frac{c|p_s|}{4\pi}+\gamma\right)+j\frac{c}{2}
\vphantom{\sum\limits_{n\ne 0}
\left[\frac{1}{jk^x_{s,l}}-\frac{c}{2\pi |l|}\right]}\right)
$$
$$
+\frac{1}{4\pi b^3} \left[ 4\sum\limits_{s=1}^{+\infty}
\frac{(2jkb+3)s+2}{s^3(s+1)(s+2)}e^{-jkbs}\cos(q_ybs) \right.
$$
$$
-(jkb+1)\left(t_+^2\log t^++t_-^2\log t^-+2e^{jkb}\cos (q_yb)\right)
$$
$$
\left. -2jkb\left(t_+\log t^++t_-\log t^-\right) +(7jkb+3)
\vphantom{\sum\limits_{s=1}^{+\infty}
\frac{(2jkb+3)s+2}{s^3(s+1)(s+2)}} \right],
$$
and the following notations are used:
$$
p_s=\sqrt{\left(k^y_s\right)^2-k^2},\qquad r_s=2q_z^2-p_s^2,
$$
$$
t^+=1-e^{-j(k+q_z)c}, \qquad t^-=1-e^{-j(k-q_z)c},
$$
$$
t_+=1-e^{j(k+q_z)c}, \qquad t_-=1-e^{j(k-q_z)c}.
$$

The calculations using \r{cfinal} can be restricted to the real part
of dynamic interaction constant $C(k,\-q,a,b,c)$  only, because its
imaginary part is given by much simpler expression (see
\cite{Belovhomo} for details): \e {\rm
Im}\left\{C(k,\-q,a,b,c)\right\}=\frac{k^3}{6\pi}. \l{imc} \f The
series in \r{cfinal} have excellent convergence that ensure very
rapid numerical calculations.

The case of magnetic scatterers can be considered using duality
principle. The expression \r{Rmain} can be used for calculation of
reflection coefficient by magnetic field (originally, this equation
represented reflection coefficient by electric field). The
dispersion equation \r{disphomo} have to be rewritten for the case
of magnetic point scatterers in the following form: \e
\mu_0\alpha_m^{-1}(\omega)=C(k,\-q,a,b,c), \l{disphomom}\f where
$\alpha_m(\omega)$ is magnetic polarizability of the scatterers. The
analytical expressions for the magnetic polarizability
$\alpha(\omega)$ of split-ring-resonators with geometry plotted in
Fig.\ref{scat} were derived and validated in \cite{SimSRR}. The
final result reads as follows \cite{Belovhomo}: \e
\alpha(\omega)=\frac{A\omega^2}{\omega_0^2-\omega^2+j\omega\Gamma},
\l{alpha} \f where $A$ is amplitude, $\omega_0$ is resonant
frequency and $\Gamma=A\omega k^3/(6\pi\mu_0)$ is the radiation
reaction factor. The expressions for amplitude $A$ and resonant
frequency $\omega_0$ in terms of dimensions of split-ring-resonators
are available in \cite{SimSRR,Belovhomo}. In the present paper we
will use typical parameters  $A=0.1\mu_0a^3$ and
$\omega_0=1/(a\sqrt{\varepsilon_0\mu_0})$.

\begin{figure}[h]
\centering \epsfig{file=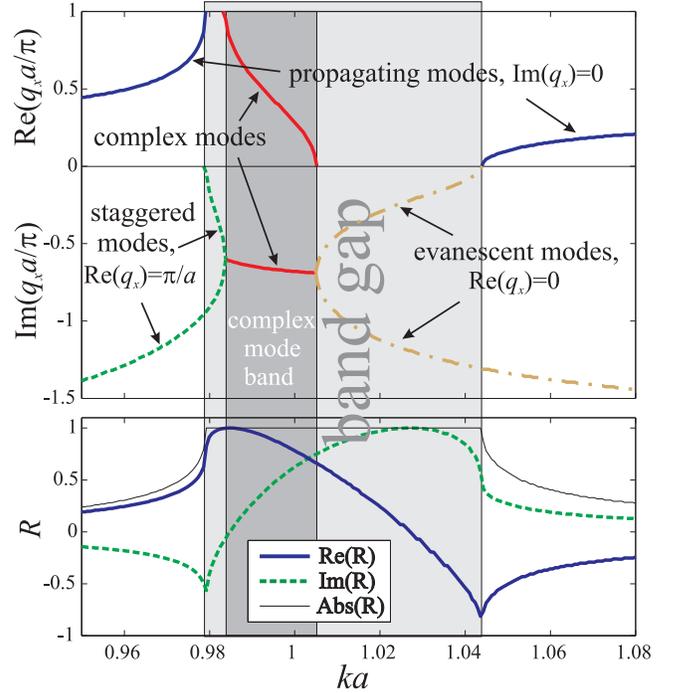, width=8.5cm} \caption{(Color
online) Dependencies of the normalized wave vectors $q_xa/\pi$ of
excited eigenmodes (imaginary and real parts) and the reflection
coefficient $R$ calculated using \r{Rmain} on normalized frequency
$ka$ for semi-infinite cubic lattice of split-ring-resonators with
$A=0.1\mu_0a^3$ and $\omega_0=1/(a\sqrt{\varepsilon_0\mu_0})$.}
\label{reflSRR}
\end{figure}

The dispersion properties of cubic lattice of split-ring-resonators
with such parameters have been extensively studied in
\cite{Belovhomo}. Using the theory of the present paper we will
study reflection properties of such a metamaterial. Let us consider
the case of cubic lattice ($a=b=c$), normal incidence ($k_y=k_z=0$)
and let the magnetic field of incident wave is along directions of
magnetic dipoles $\vec d=\vec y_0$. Numerical solution of dispersion
equation \r{disphomom} with $q_y=k_y=0$ and $q_z=k_z=0$ allows to
get a set of wave vectors of excited eigenmodes $\{q^x_i\}$. These
wave vectors are plotted at the top of Fig. \ref{reflSRR} as
functions of normalized frequency $ka$. The point $ka=1$ corresponds
to the resonant frequency $\omega_0$ os split-ring-resonators. One
can see that the propagating modes (${\rm Im}(q_x)=0$) exist only
for $ka\le 0.978$ and $ka\ge 1.044$. It means that a partial
resonant band gap is observed for $ka\in [0.978,1.044]$. At the
frequencies inside of the band gap all the eigenmodes decay with
distance. Note, that such decaying modes exist at all frequencies,
not only inside of the band gap. The Fig. \ref{reflSRR} shows only
the eigenmodes with slowest decay $|{\rm Im}(q_x)|<1.5\pi/a$. There
is an infinite number of other decaying modes which decay with
distance more rapidly. The contribution of such modes into the
reflection coefficient is negligible as it was shown above. The
decaying modes can be separated into the three classes:
\begin{itemize}
\item evanescent modes, the modes which have ${\rm Re}(q_x)=0$, they
decay exponentially from one crystal plane to the other one;

\item staggered modes,  the modes which have ${\rm Re}(q_x)=\pi/a$,
they exponentially decay from one crystal plane to the other one by
absolute value, but the dipoles in the neighboring plane are excited
in out of phase;

\item complex modes, the most general case of the decaying modes which have ${\rm Re}(q_x)\ne
0$, they experience both exponential decay and phase variation from
one crystal plane to the other one.
\end{itemize}

The evanescent modes are the most common type of decaying modes.
They can be observed in dielectrics with negative permittivity, for
example. The staggered modes are limiting case of the complex modes
and can be widely observed in periodical structures in vicinity of
the band gap edges, see for example \cite{Belovnonres,lwPRE}. The
complex modes of general kind are quite exotic for common materials.

In the system under consideration we are able to observe all three
kind of mentioned decaying modes. The presence of staggered and
complex modes are evidence of spatial dispersion in this material
reported in \cite{Belovhomo}. The staggered modes exists for $ka\ge
0.984$, evanescent modes for $ka\ge 1.015$ and complex modes for
$ka\in [0.984,1.015]$ (see Fig. \ref{reflSRR}). One can see that for
a fixed frequency from the range $ka\in [0.978,0.984]$  there are
two staggered modes and in the range $ka\in [1.015,1.044]$ there are
two evanescent modes. Actually, in the range $ka\in [0.984,1.015]$
there are also two complex modes which have the same imaginary parts
but differs by sign of the real part. Thus, we conclude that for
every frequency from the range $ka\in [0.95,1.08]$ that we consider
the incident wave will excite in the crystal a pair of modes with
$|{\rm Im}(q_x)|<1.5\pi/a$ (propagating and staggered, two
staggered, two complex, two evanescent or propagating and
evanescent). Using the usual approach one has to introduce an
additional boundary condition to solve this problem since usual
condition of tangential component continuity is not enough in the
case of excitation of two modes.

Using the theory introduced in the previous section it is enough to
substitute obtained wave vectors of eigenmodes into \r{Rmain} the
reflection coefficient is calculated. The reflection coefficient is
plotted at the bottom of Fig. \ref{reflSRR}. One can see that at the
frequencies in vicinity of the bottom and top edges of the band gap
the semi-infinite crystal operate nearly as the electric and
magnetic walls, respectively, as it was predicted in
\cite{Belovnonres}. At the frequency $ka=0.984$ the reflection
coefficient is equal to $+1$ (electric wall), and at $ka=1.044$ it
is $-0.8+0.6j$ (nearly magnetic wall). Note, that the frequency
corresponding to the electric wall effect is not equal to the bottom
edge of the band gap and there are no frequency exactly
corresponding to the magnetic wall effect. The use of the usual
formulae for reflection coefficient from magnetic and
Clausius-Mossotti formulae which do not take into account effects of
spatial dispersion one could get idea that magnetic and magnetic
wall effects have to happen at the edges of the band gap. Our study
demonstrate that if the spatial dispersion is taken into account
accurately then it is not so.

Thus, we have demonstrated how the proposed theory can be used for
modeling of reflection from semi-infinite crystals with spatial
dispersion. Our theory can be treated as generalization of results
of Mahan and Obermair \cite{Mahan} which have been widely applied
for modeling of various kinds of reflection problems. We hope that
the present generalization can find much more applications in
modeling of reflection  from spatially dispersive materials since in
has no restriction on the period of the lattice to be smaller than
wavelength and allows to consider electromagnetic crsytals of
general kind.

\section{Conclusion}

In this paper a new approach for solving problems of plane-wave
diffraction on semi-infinite electromagnetic crystals is proposed.
The boundary conditions for the interface between isotropic
dielectric and electromagnetic crystal of general kind are deduced
in the form of infinite system of equations relating amplitudes of
incident wave, excited eigenmodes and scattered spatial harmonics.
This system of equations represent mathematical content of
generalized Ewald-Oseen extinction principle which is formulated in
this paper: the polarization of the semi-crystal excited by plane
wave is distributed in such way that it cancels the incident wave
together with all high-order spatial harmonics associated with
periodicity of the boundary. In our opinion, the proof of
generalized Ewald-Oseen extinction principle presented in this paper
is an important theoretical fact which helps to understand
interrelation between reflection and dispersion properties of
electromagnetic crystals. If the eigenmodes of the infinite crystal
are known then the system can be solved numerically which provides
new numerical method for solving the diffraction problem under
consideration. We believe in the quite good prospects for the
application of the described method in further studies of dielectric
and even metallic electromagnetic crystals at both microwave and
optical ranges.

For the special case when the crystal is formed by small scatterers
which can be effectively replaced by dipoles with fixed orientation
the deduced system of equations is solved analytically using method
of the characteristic function. The closed form expressions for the
amplitudes of excited eigenmodes and scattered spatial harmonics are
provided in terms of rapidly convergent products. These expressions
can be treated as generalization of classical result of Mahan and
Obermair \cite{Mahan} for the case when period of the lattice can be
large as compared to the wavelength. The proposed method is applied
for calculation of reflection coefficient from semi-infinite crystal
formed by resonant magnetic scatterers (split-ring-resonators) at
the frequencies corresponding to the strong spatial dispersion.

\section*{Acknowledgements}
The authors would like to thank Sergei Tretyakov, Stanislav
Maslovski and  Mario Silveirinha for useful discussions of the
material presented in this paper.

\bibliography{abc}
\end{document}